\documentclass[12pt]{iopart}
\usepackage[dvips]{graphicx}

\begin{document}

\title[Quantum lattice solitons]
{Quantum lattice solitons in ultracold bosons near the Feshbach resonance}

\author{K V Krutitsky$^1$ and D V Skryabin$^2$}

\address
{$^1$
Fachbereich Physik der Universit\"at Duisburg-Essen, Campus Essen,
Universit\"atsstr.~5, 45117 Essen, Germany
}
\address
{$^2$
Department of Physics, University of Bath, Bath, BA2~7AY, UK
}
\ead{kostya@theo-phys.uni-essen.de}

\begin{abstract}
Quantum lattice solitons in a system of two
ultracold bosons near Feshbach resonance are investigated.
It is shown that their binding energy,
effective mass, and spatial width, can be manipulated varying the detuning from
the Feshbach resonance.
In the case of attractive atomic interactions, the molecule creation stabilizes
the solitons.
In the case of repulsive interactions, the molecule creation leads
to the possibility of existence of bright solitons in some interval of detunings.
Due to quantum fluctuations the distance between the atoms is a random quantity
with the standard deviation larger than the mean value.
\end{abstract}



\maketitle

\section{Introduction}

Many-body phenomena in ultracold quantum gases is a subject of extensive
ongoing research. Interaction between atoms plays a
crucial role in many situations and is responsible for the most
striking experimental observations of solitons in
Bose-Einstein condensates trapped by harmonic~\cite{becsol} and periodic~\cite{markus}
potentials. The typical number of atoms in such solitons
varies from several hundreds to tens of thousands.
Therefore these structures can be well described by the mean-field Gross-Pitaevsky
equation. However, the mean-field theory does never provide an exact description
of interacting quantum systems (e.g., due to unavoidable depletion) and it becomes
interesting to investigate quantum effects in the soliton propagation~\cite{bth}.
If the atoms are loaded into the optical lattice, the interaction effects become more
important and in the limit of the small number of atoms an exact quantum analysis 
of the solitons~\cite{Scott,el}
reveals strong deviations from the results provided by the
Gross-Pitaevsky equation with lattice potential~\cite{skryabin} or
its discrete version~\cite{Scott,discr}.

The use of Feshbach resonances to control interaction between
ultracold atoms in optical potentials is a widely spread technique allowing
transformation of atoms into molecules and changing  magnitude and
sign of the effective scattering length of the atoms (see, e.g.,~\cite{MVA,fesh}).
Coherent solitons in condensed atomic-molecular mixtures
without optical lattice were studied in several papers, see,
e.g.,~\cite{par1,par2}. The mathematical model previously used for the coupled
atomic-molecular condensates is equivalent to the model describing
parametric interaction of photons in quadratically nonlinear
crystals~\cite{rev}. It was demonstrated in the mean-field
limit that the resonant atomic-molecular interaction serves as a
mechanism responsible for supporting bright solitons in the case
of repulsive bosons and for preventing collapse in the case of
attractive bosons~\cite{par2,rev}. The quantum atomic-molecular
solitons in the system without periodic potential were also
studied~\cite{par1,quant}.
Optical parametric solitons in
the system of coupled waveguides, playing the role of a periodic potential
for photons, were recently observed experimentally~\cite{steg}.
Atomic-molecular solitons in a
deep optical lattice have been theoretically considered in the
mean-field approximation~\cite{konotop}, which is mathematically equivalent
to the system studied in~\cite{steg}.

In this work, we demonstrate existence and study
properties of the quantum atomic-molecular solitons in an optical
lattice near Feshbach resonance.
Under the quantum lattice soliton we understand the quantum state of the system
of interacting particles with the localized eigenfunction and the discrete
energy level belonging to a spectral interval forbidden for the spatially extended
periodic states~\cite{Scott}.
Note that the discrete energy levels belonging to the intervals forbidden
for the linear waves are also a generic feature of the classical lattice solitons.
Advances in manipulation of ultracold atomic systems with small number of particles per
lattice site~\cite{Greiner} as well as in cooling and trapping of single
atoms~\cite{single} allow one to hope that quantum lattice
solitons will soon become relevant for experimental research.

\section{Hamiltonian}

We consider two atoms of mass $m$ in an optical lattice created by a
far-detuned standing laser wave. If the laser wavelength is $\lambda_{\rm L}=2\pi/k_{\rm L}$,
then the lattice constant $d=\lambda_{\rm L}/2$. It is convenient
to represent the amplitude of the periodic potential in the form $\hbar\omega_{\rm R} s$,
where $\omega_{\rm R}=\hbar k_{\rm L}^2/(2 m)$ is the recoil frequency and
$s$ is a dimensionless parameter.
In the case of a deep optical lattice every lattice site can be described by a harmonic potential
with the frequency $\omega=2\omega_{\rm R}\sqrt{s}$
and the lowest-band atomic Wannier function
is well approximated by a Gaussian with the characteristic length
$l_{\rm a}=\sqrt{\hbar/m \omega}$.
The atoms in the lattice are subject to the magnetic field $B$,
with $B=B_0$ corresponding to the Feshbach resonance of the width $\Delta B$.

There are several processes which are to be
taken into account in such a system: atomic interaction, molecule production and
atomic and molecular hopping.
Taking into account only the hopping between the nearest lattice sites
as well as on-site atomic interactions and in the lowest-band approximation
the Hamiltonian of the system is given by~\cite{DKOS,comment}
\begin{eqnarray}
\label{bh}
H
&=&
-
t_{\rm a}
\sum_{\langle i,j \rangle}
a^{\dagger}_{i}
a^{\phantom \dagger}_{j}
-
t_{\rm m}
\sum_{\langle i,j \rangle}
b^{\dagger}_{i}
b^{\phantom \dagger}_{j}
+
\left(
    \delta-\frac{3}{2}\hbar\omega
\right)
\sum_{i}
b^{\dagger}_{i}
b^{\phantom \dagger}_{i}
+
\frac{U_{\rm bg}}{2}
\sum_{i}
a^{\dagger}_{i}
a^{\dagger}_{i}
a^{\phantom \dagger}_{i}
a^{\phantom \dagger}_{i}
\nonumber\\
&+&
\tilde g
\sum_{i}
\left(
b^{\dagger}_{i}
a^{\phantom \dagger}_{i}
a^{\phantom \dagger}_{i}
+
a^{\dagger}_{i}
a^{\dagger}_{i}
b^{\phantom \dagger}_{i}
\right)
\;,
\end{eqnarray}
where
$a^{\dagger}_{i}$~($b^{\dagger}_{i}$)
and
$a^{\phantom \dagger}_{i}$~($b^{\phantom \dagger}_{i}$)
are creation and annihilation operators of a single atom (molecule) at a lattice site $i$,
$\delta=\Delta\mu(B-B_0)$ is a detuning from the Feshbach resonance.
Here, $\Delta\mu$ is the difference in magnetic moments of the two atoms and a molecule.
The atom-molecule conversion is determined by
$
\tilde g
=
\hbar
\sqrt{2 \pi a_{\rm bg}
\Delta B \Delta \mu /m}
/(2 \pi l_{\rm a}^2)^{3/4}
$
and the background on-site atomic interaction parameter is
$
U_{\rm bg}
=
\sqrt{2/\pi}
\hbar \omega
\left(
    a_{\rm bg} / l_{\rm a}
\right)
$
with $a_{\rm bg}$ being the background scattering length.
In the Gaussian approximation, the atomic and molecular tunneling matrix elements are given by
$
t_{\rm a,m}
=
\frac{\hbar \omega}{2}
\left[
    1 - \left( \frac{2}{\pi} \right)^{2}
\right]
\left(
    \frac{\lambda_{\rm L}}{4 l_{\rm a,m}}
\right)^{2}
e^{
    -
    \left(
        \lambda_{\rm L}/4 l_{\rm a,m}
    \right)^{2}
  }
$.
Since
$l_{\rm m} = l_{\rm a}/\sqrt{2}$,
the molecular tunneling rate is much smaller than the atomic one.

\section{Solution of the on-site problem}

The on-site problem for the Hamiltonian (\ref{bh}) can be easily solved
analitically. In the case when the atoms are on the same lattice site
there are two eigenmodes which are superpositions of the two-atom and
molecular states with the energies
\begin{equation}
\label{e}
E_\pm
=
\frac
{\delta'+U_{\rm bg}}
{2}
\pm
\sqrt
{
  \left(
      \frac
      {\delta'-U_{\rm bg}}
      {2}
  \right)^2
  +
  2
  \tilde g^2
}
\;,
\end{equation}
and the probability to find a molecule
\begin{equation}
\label{pm}
p_{{\rm m}\pm}
=
\frac{1}{2}
\left[
    1
    \pm
    \frac
    {\delta'-U_{\rm bg}}
    {
      \sqrt
      {
       \left(
           \delta'-U_{\rm bg}
       \right)^2
       +
       8
       \tilde g^2
      }
    }
\right]
\;,
\end{equation}
where
$\delta'=\delta-\frac{3}{2}\hbar\omega$
is an effective detuning.

\begin{figure}[tb]
\centering


\hspace{-3cm}
  \includegraphics[width=8cm]{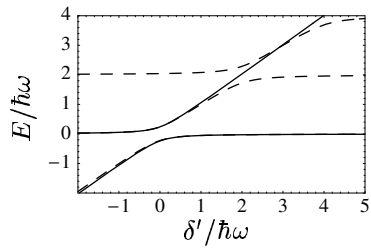}

\caption{Eigenenergies in the case of two atoms on the same lattice site.
         Solid lines show the results given by Eq.~(\ref{e}) which corespond
	 to the lowest-band approximation. The results for the infinite number of bands [Eq.~(\ref{emo})]
	 are shown by dashed lines.
	 $U_{\rm bg}=0$,
	 $2 \sqrt{\pi} \tilde g^2 / \left(\hbar\omega\right)^2=0.1$.
	}
\label{ee}
\end{figure}

The two-atoms on-site problem was exactly solved in Ref.~\cite{DKOS} for the infinite number of bands
neglecting the atom-atom interaction.
The eigenenergies $E$ are shown to be determined by the equation
\begin{equation}
\label{emo}
E - \delta'
=
\frac{2\sqrt{\pi}\tilde g^2}{\hbar \omega}
\frac
{\Gamma (- E / 2 \hbar \omega)}
{\Gamma (- E / 2 \hbar \omega - 1/2)}
\;.
\end{equation}
The eigenenergies given by Eq.~(\ref{e}) for $U_{\rm bg}=0$ and Eq.~(\ref{emo})
are plotted in Fig.~\ref{ee}. As we see, our lower-branch solution $E_-$ in Eq.~(\ref{e})
is in excellent agreement with the corresponding branch of Eq.~(\ref{emo}) for arbitrary $\delta$.
The upper-branch solution $E_+$ fails to reproduce the second branch of Eq.~(\ref{emo})
if $\delta$ is far above the Feshbach resonance where the contribution of the second band
becomes significant, remaning however in a very good agreement near the resonance
and below it. This implies
that the lowest-band approximation is valid if the effective detuning
$\delta'$ is less than
the gap between the two lowest Bloch bands, which is the quantity of the order of $\hbar\omega$,
and/or if we are interested in the eigenmodes of the Hamiltonian (\ref{bh})
with the energies less than the energy of the second Bloch band.
The latter is always the case in the present work.
In addition, the parameters $U_{\rm bg}$ and $\tilde g$ must be much smaller than
the bands separation which is also fulfilled.

\section{Eigenmodes of the complete Hamiltonian and the soliton band}

We consider a one-dimensional model with $L$ lattice sites
and assume that $L$ is odd\footnote{In the case of even $L$ there will be only unessenstial modifications
in the equations.}.
Under periodic boundary conditions the eigenstates of the Hamiltonian~(\ref{bh}) are
\begin{eqnarray}
\label{psi}
|\psi_k\rangle
&=&
c^{\rm m}_{k}
\sum_{j=1}^L
\left(
    \hat T/\tau_k
\right)^{j-1}
|1_{\rm m} 0 \dots 0\rangle
+
c^{\rm a}_{0k}
\sum_{j=1}^L
\left(
    \hat T/\tau_k
\right)^{j-1}
|2 0 \dots 0\rangle
\nonumber\\
&+&
c^{\rm a}_{1k}
\sum_{j=1}^L
\left(
    \hat T/\tau_k
\right)^{j-1}
|1 1 0 \dots 0\rangle
+
\dots
\nonumber\\
&+&
c^{\rm a}_{(L-1)/2,k}
\sum_{j=1}^L
\left(
    \hat T/\tau_k
\right)^{j-1}
|1 0 \dots 0 1 0 \dots 0\rangle
\;,
\end{eqnarray}
where $|1_{\rm m} 0 \dots 0\rangle$ is a state with one molecule
on the first lattice site and all the other sites being unoccupied,
$|n_1 \dots n_L\rangle$ is a state with $n_i$ atoms on site $i$, $i=1,\dots,L$.
$\hat T$ is the translation operator which has the eigenvalues
$
 \tau_k
 =
 \exp
 \left(
     i \pi k/k_{\rm L}
 \right)
$
with the wave number
$k=k_{\rm L} 2 \nu/L$, $\nu=0,\pm 1,\dots,\pm (L-1)/2$~\cite{Scott}.
The eigenvalue problem for the Hamiltonian~(\ref{bh}) can be written down in the matrix form
\begin{equation}
\label{evp}
\left(
    \begin{array}{cc}
       \epsilon_k^{\rm m} & A^T \\
       A & Q_k
    \end{array}
\right)
\left(
    \begin{array}{c}
       {c}_k^{\rm m} \\
       {\bf c}_k^{\rm a}
    \end{array}
\right)
=
E_k
\left(
    \begin{array}{c}
       {c}_k^{\rm m} \\
       {\bf c}_k^{\rm a}
    \end{array}
\right)
\;,
\end{equation}
where
$
 \epsilon_{k}^{\rm m}
 =
 \delta'
 -
 2 t_{\rm m}
 \cos
 \left(
     \pi k/k_{\rm L}
 \right)
$.
The vector $A$ has a length $(L+1)/2$ and its nonvanishing
element is $A_{1}=\sqrt{2}\tilde g$.
The nonvanishing elements of the tridiagonal $(L+1)/2 \times (L+1)/2$ matrix $Q_k$
are given by~\cite{Scott}
\begin{eqnarray}
&&
Q_{11}=U_{\rm bg}
\;,\;
Q_{21} = Q_{12}^* = - t_{\rm a} \sqrt{2} (1+\tau_k)
\;,
\\
&&
Q_{i+1,i} = Q_{i,i+1}^* = - t_{\rm a} (1+\tau_k)
\;,\;
i=2,\dots,(L-1)/2
\;,
\nonumber\\
&&
Q_{(L+1)/2,(L+1)/2}
=
- t_{\rm a}
\left[
    \tau_k^{(L+1)/2}
    +
    \tau_k^{(L-1)/2}
\right]
\;.
\nonumber
\end{eqnarray}
The eigenvectors in Eq.~(\ref{evp}) consist of two parts
${c}_k^{\rm m}$,
$
{\bf c}_k^{\rm a}
=
{\rm col}
\left[
    c^{\rm a}_{0k},\dots,c^{\rm a}_{(L-1)/2,k}
\right]
$,
and satisfy the normalization condition
\begin{eqnarray}
\label{norm}
\left|
    c_{k}^{\rm m}
\right|^2
+
\sum_{i=0}^{(L-1)/2}
\left|
    c_{ik}^{\rm a}
\right|^2
&=&
1
\;.
\end{eqnarray}

In the absence of the molecular mode, the eigenvalue problem~(\ref{evp}) reduces
to that one solved in Ref.~\cite{Scott}, where it was shown that in the case of attractive
interaction the energy spectrum consists always of (quasi)continuum band
and a discrete level below the (quasi)continuum which corresponds to the bright soliton.
Its characteristic feature is that
$
  \left|
      c_{0k}^{\rm a}
  \right|^2
  \gg
  \left|
      c_{ik}^{\rm a}
  \right|^2
$,
$i=1,\dots,(L-1)/2$, i.e., the probability of finding two atoms on the same lattice site
is much higher than all the other ones.
This localization corresponds to the soliton solution of the discrete nonlinear
Schr\"odinger equation and, therefore, the discrete level can be called
a "soliton band"~\cite{Scott}. Our aim is to investigate
the influence of the molecular mode on the soliton band.

After the eigenvalue problem~(\ref{evp}) is solved, one can calculate
the soliton binding energy $E_{\rm b}$
which is defined as the difference of the energy at the bottom
of the (quasi)continuum and the soliton level at $\nu=0$ which corresponds to $k=k_0=0$.
The effective mass $m^*$ can be worked out using a quadratic approximation for the eigenenergy
at some small value of $\nu$ (e.g., $\nu=1$)
$
E_{k_1}
=
E_{k_0}
+
\hbar^2 k_1^2
/
\left(
    2 m^*
\right)
$,
which leads to
\begin{eqnarray}
m^*
&=&
 2 \hbar^2 k_{\rm L}^2
 /
\left[
 \left(
     E_{k_1}
     -
     E_{k_0}
 \right)
 L^2
\right]
\;.
\end{eqnarray}
According to Eq.~(\ref{psi}) the distance between the atoms $w_k$
is a random variable which takes the values
$w_{ki}=0,1,\dots,(L-1)/2$,
with the probabilities
$
\left|
     c_{ik}^{\rm a}
\right|^2
/
\left(
    1
    -
    \left|
        c_{k}^{\rm m}
    \right|^2
\right)
$.
Thus, it is necessary to calculate
not only the mean interatomic distance $\langle w_k \rangle$ but also its standard deviation
\begin{eqnarray}
\sigma_{wk}
&=&
\sqrt
{
\langle w_k^2 \rangle
-
\langle w_k \rangle^2
}
\;,
\langle w_k^l \rangle
=
\sum_{i=0}^{(L-1)/2}
\frac
{
 i^l
 \left|
     c_{ik}^{\rm a}
 \right|^2
}
{
 1
 -
 \left|
     c_{k}^{\rm m}
 \right|^2
}
\;,
\end{eqnarray}
and the soliton width can be defined as
$
\sqrt
{
 \langle w_k^2 \rangle
}
$.

We have solved the eigenvalue problem~(\ref{evp}) numerically for finite values of $L$
and analytically in the limit of infinite lattice.
The results are presented below.
We consider the cases of attractive and repulsive atomic interactions and
concentrate on the properties of the lower-energy modes.

\subsection{Attractive atomic interaction}

\begin{figure}[tb]
\centering


\hspace{-3cm}
  \includegraphics[width=8cm]{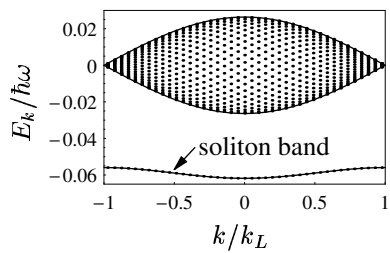}

\caption{Energy eigenvalues of the Hamiltonian~(\ref{bh}).
         The parameters are $s=5$,
         $2 \sqrt{\pi} \tilde g^2 / \left(\hbar\omega\right)^2=0.1$,
         $\delta'/\hbar\omega=3$,
	 $a_{\rm bg}/\lambda_L=-0.005$.
	 Dots are the results of numerical solution of Eq.~(\ref{evp}) for $L=41$ and
	 the solid lines correspond to the limit $L\to\infty$.
	 The spectrum is truncated from above in order to be consistent
	 with the lowest-band approximation.
	}
\label{s}
\end{figure}

\begin{figure}[tb]
\centering



\hspace{-3cm}
  \includegraphics[width=8cm]{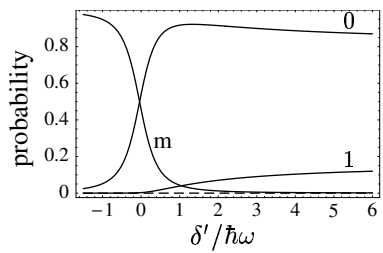}

\caption{Probabilities of molecular and atomic states corresponding to the soliton band:
         $
	  \left|
	      c_{0}^{\rm m}
	  \right|^2
	 $ (m),
	 $
	  \left|
	      c_{i0}^{\rm a}
	  \right|^2
	 $ ($i$),
	 $i=0,1,2$
	 [$
	   \left|
	       c_{20}^{\rm a}
	   \right|^2
	  $
	  is shown by the dashed line].
         The parameters are the same as in Fig.~\ref{s}.
	}
\label{pn}
\end{figure}

\begin{figure}[tb]
\centering





\hspace{-3cm}
  \includegraphics[width=8cm]{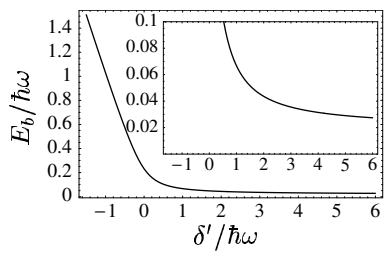}

\caption{Soliton binding energy.
         The parameters are the same as in Fig.~\ref{s} and $k=0$.
	}
\label{ebn}
\end{figure}

\begin{figure}[tb]
\centering





\hspace{-3cm}
  \includegraphics[width=8cm]{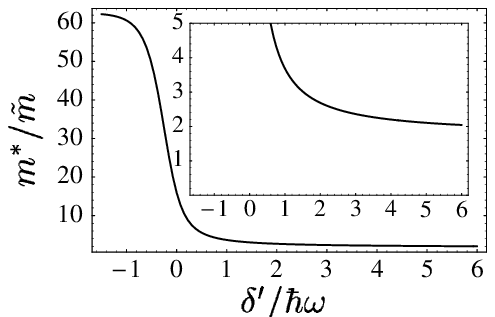}

\caption{The ratio of the soliton effective mass $m^*$ to the effective mass
         $\tilde m$ at the bottom of (quasi)continuum band.
         The parameters are the same as in Fig.~\ref{s}.
	}
\label{mn}
\end{figure}

\begin{figure}[tb]
\centering


\hspace{-3cm}
  \includegraphics[width=10cm]{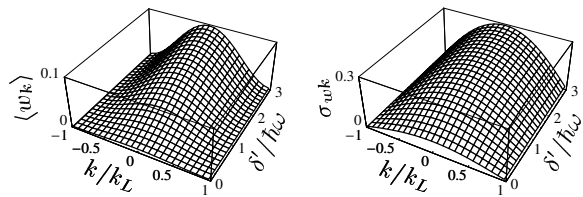}

\caption{Mean interatomic distance $\langle w_k \rangle$ (left panel)
         and its standard deviation $\sigma_{wk}$ (right panel).
         The parameters are the same as in Fig.~\ref{s}.
	}
\label{wdn}
\end{figure}

We consider first bosons with attractive interactions ($U_{\rm bg}<0$).
If $\delta'$ is negative and its absolute value is very large,
the coupling between the molecular
mode and the atomic mode is negligible and we have two discrete levels below the (quasi)continuum.
The lower one corresponds to the pure molecule and another one to the atomic bright soliton.
If $\delta'$ increases, i.e, we come closer to the Feshbach resonance,
both discrete levels approach the (quasi)continuum.
At some critical value of $\delta'=\delta'_-$
the upper level merges with the (quasi)continuum.
In the numerical calculations it is not quite clear how to determine $\delta'_-$
exactly because there are several possibilities to define it.
Analytical analysis in the case of the infinite lattice shows that the mergence occurs
if at least one of the inequalities
\begin{equation}
\label{ineq}
\left|
    c_{1k}^{\rm a}
\right|
>
\left|
    c_{ik}^{\rm a}
\right|
\;,
i=2,3,\dots,
\end{equation}
is violated. We adopt this as a definition of $\delta'_-$ and by doing numerical diagonalization
for $k=0$ and for the values of parameters in the caption of Fig.~\ref{s} we obtain
$\delta'_-=-1.531\,\hbar\omega$. In order to have inequalities (\ref{ineq}) again fulfilled,
one has to increase $\delta'$ up to $\delta'_+$. Using the same values of the parameters
we get $\delta'_+=-1.490\,\hbar\omega$. If $\delta'>\delta'_+$, a discrete level
appears above the (quasi)continuum, while the lower one
which becomes a linear combination of atomic and molecular states remains below (see Fig.~\ref{s}).
If we increase $\delta'$ further and go far away from the Feshbach resonance
($\delta' \gg \hbar\omega$),
the contribution of the molecular mode into the lowest-energy eigenstate becomes negligible~(Fig.\ref{pn})
and we have a pure atomic bright soliton below the (quasi)continuum~\cite{Scott}.
The upper discrete level is located very far above the (quasi)continuum and cannot be interpreted
within the lowest-band approximation.

The soliton binding energy $E_{\rm b}$ is shown in Fig.~\ref{ebn}.
Due to the large contribution of the molecular mode
near the resonance the binding energy is larger than its asymptotic value at $\delta'\to\infty$.
The effective mass $m^*$ is also larger at smaller values of $\delta'$ (see Fig.~\ref{mn})
because due to the fact that $t_{\rm m} \ll t_{\rm a}$
the effective mass of the molecule is much larger than the atomic effective mass.
The corresponding contributions of the molecular and atomic states into the soliton band
are shown in Fig.~\ref{pn}.

The mean interatomic distance $\langle w_k \rangle$ as well as its standard deviation $\sigma_{wk}$
are shown in Fig.~\ref{wdn}. The interatomic distance is well below the lattice constant $d$
and the maximal localization is achieved at the edges of the Brillouin zone.
However, quantum fluctuations are very strong and $\sigma_{wk}>\langle w_k \rangle$.

\subsection{Repulsive atomic interaction}

\begin{figure}[tb]
\centering



\hspace{-3cm}
  \includegraphics[width=8cm]{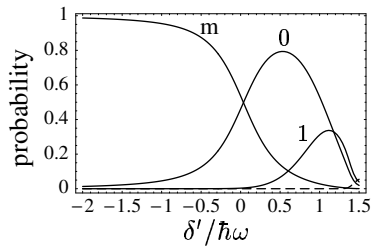}

\caption{Probabilities of molecular and atomic states corresponding to the soliton band:
         $
	  \left|
	      c_{0}^{\rm m}
	  \right|^2
	 $ (m),
	 $
	  \left|
	      c_{i0}^{\rm a}
	  \right|^2
	 $ ($i$),
	 $i=0,1,2$
	 [$
	   \left|
	       c_{20}^{\rm a}
	   \right|^2
	  $
	  is shown by the dashed line].
	 $a_{\rm bg}/\lambda_L=0.005$ and the other
         parameters are the same as in Fig.~\ref{s}.
	}
\label{pp}
\end{figure}

\begin{figure}[tb]
\centering


\hspace{-3cm}
  \includegraphics[width=8cm]{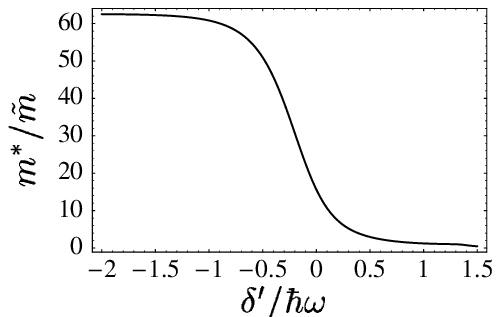}

\caption{The ratio of the soliton effective mass $m^*$ to the effective mass
         $\tilde m$ at the bottom of (quasi)continuum band.
         The parameters are the same as in Fig.~\ref{pp}.
	}
\label{mp}
\end{figure}

\begin{figure}[tb]
\centering


\hspace{-3cm}
  \includegraphics[width=10cm]{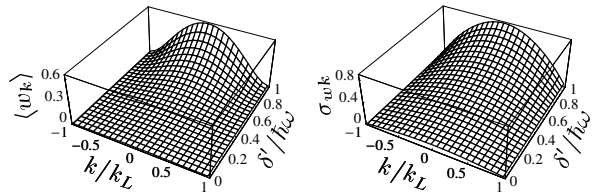}

\caption{Mean interatomic distance $\langle w_k \rangle$ (left panel)
         and its standard deviation $\sigma_{wk}$ (right panel).
         The parameters are the same as in Fig.~\ref{pp}.
	}
\label{wdp}
\end{figure}

In the case of repulsive atomic interaction ($U_{\rm bg}>0$), the situation is quite different.
There can be only one discrete level below the (quasi)continuum which is occupied
by the molecule as long as $\delta'$ is negative and its absolute value remains very large.
Above the (quasi)continuum, there is another discrete level corresponding to the atomic
bright soliton. If we increase $\delta'$
the lower discrete level approaches the (quasi)continuum and the probabilities of the atomic
states become larger meaning that the system enters the bright soliton regime
supported by the molecule creation (see Fig.~\ref{pp}).
Inequalities (\ref{ineq}) are satisfied in this regime.
If we increase $\delta'$ further and reach the value $\delta'_-$,
the probability of the molecular state becomes very small.
Inequalities (\ref{ineq}) are violated and
the discrete level merges with the (quasi)continuum, i.e.,
the bright soliton is destroyed.
The probability
$
\left|
    c_{k}^{\rm m}
\right|^2
$
can be also interpreted as a relative population of the molecular component.
It is a decreasing function of $\delta'$ like in the case of classical
atomic-molecular solitons~\cite{konotop}.

If the detuning is further increased up to $\delta'_+$,
the soliton band appears above the (quasi)continuum.
For the values of parameters used in our numerical estimations,
$\delta_-'=1.479\,\hbar\omega$ and
$\delta_+'=1.521\,\hbar\omega$.

The soliton binding energy $E_{\rm b}$ is again
a decreasing function of $\delta'$ which vanishes at $\delta'=\delta_-'$.
The effective mass $m^*$ equals to the effective mass of the molecule for large
negative $\delta'$ and reaches the value $\tilde m$ at $\delta'=\delta_-'$ (Fig.~\ref{mp}).
If we come closer to $\delta_-'$ the solitons become less localized especially at $k=0$
and the interatomic-distance fluctuations increase (Fig.~\ref{wdp}).
According to our definition of the soliton width, its behavior is similar to that of
$\langle w_k \rangle$ and $\sigma_{wk}$ shown in Fig.~\ref{wdp}.

\subsection{The limit $L\to\infty$}

\begin{figure}[tb]
\centering

%
%
\hspace{-3cm}
  \includegraphics[width=8cm]{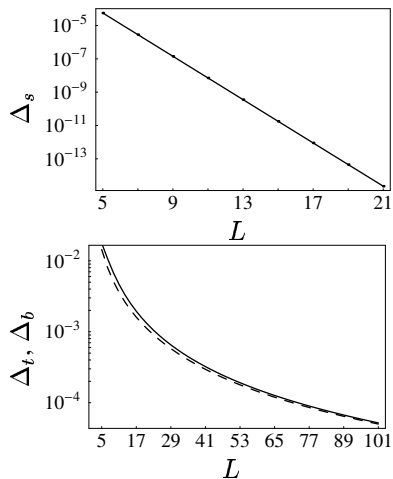}

\caption{Deviations of the eigenvalues of Eq.(\ref{evp}) for finite $L$ from that
         obtained in the limit $L\to\infty$.
	 $\Delta_t$ is shown by the dashed line.
	 The parameters are the same as in Fig.~\ref{s}.
	}
\label{d}
\end{figure}

In the limit $L\to\infty$, the wave number $k$ becomes a continuous variable.
The energies of the continuum band are enclosed in the interval
$
 \left|
     E_k
 \right|
 \le
 q_k
$,
where
$
 q_k
 =
 4 t_{\rm a}
 \cos
 \left(
     \frac{\pi}{2}
     \frac{k}{k_{\rm L}}
 \right)
$,
and the coefficients
$c_{0k}^{\rm a}$, $c_{k}^{\rm m}$,
in Eq.~(\ref{evp}) become negligibly small.
Outside of the continuum band, the solutions have the form
$
 c_{jk}^{\rm a}
 =
 a_k b_k^j
 \exp
 \left(
     i
     \frac{\pi}{2}
     \frac{k}{k_{\rm L}}
     j
 \right)
$,
$j=1,2,\dots,\infty$.
Substituting this ansatz into Eq.~(\ref{evp}) we obtain the equation
for the eigenenergy ${\cal E}_k = \lim_{L\to\infty}E_k$
\begin{eqnarray}
\label{E}
{\cal E}_k^2
&=&
\left(
    U_{\rm bg}
    +
    U_k
\right)^2
+
q_k^2
\;,\;
U_k
=
2
\tilde g^2
/
\left(
    {\cal E}_k-\epsilon_{k}^{\rm m}
\right)
\;.
\end{eqnarray}
Note that the quantity $U_{\rm bg} + U_k$ plays a role of the effective atomic
interaction.
The values of $a_k$ and $b_k$ corresponding to a certain ${\cal E}_k$ are given by
$
a_k
=
\sqrt{2}
c_{0k}^{\rm a}
$,
$
b_k
=
\left(
    U_{\rm bg}
    +
    U_k
    -
    {\cal E}_k
\right)/q_k
$,
and the expressions for the probabilities of the state with two atoms
on the same lattice site and the molecular state take the form
\begin{eqnarray}
\label{c0a}
\left|
    c_{0k}^{\rm a}
\right|^2
&=&
\left(
    1-b_k^2
\right)
/
\left[
 1
 +
 b_k^2
 +
 \left(
     1 - b_k^2
 \right)
 S_k
\right]
\;,
\nonumber\\
\left|
    c_{k}^{\rm m}
\right|^2
&=&
2
\tilde g^2
\left|
    c_{0k}^{\rm a}
\right|^2
/
 \left(
     {\cal E}_k
     -
     \epsilon_{k}^{\rm m}
 \right)^2
\;,
\end{eqnarray}
where
$
S_k
=
2
\tilde g^2
/
{
\left(
    {\cal E}_k
    -
    \epsilon_{k}^{\rm m}
\right)^2
}
$.

Eq.~(\ref{E}) can be multiplied by
$
 \left(
     {\cal E}_k
     -
     \epsilon_{k}^{\rm m}
 \right)^2
$
and treated as quartic equation for ${\cal E}_k$ which contains always four roots.
However, depending on the values of the parameters only one or two roots are real
and provide normalized eigenstates implying that the others are unphysical
and should be rejected. The normalization condition (\ref{norm}) requires
$
 a_k^2/2<1
$
as well as
$
  \left|
      b_k
  \right|
  < 1
$.
One can easily show that in the special case $t_{\rm a}=t_{\rm m}=0$ the physical solutions
of Eq.~(\ref{E}) are given by (\ref{e}).

We substitute ${\cal E}_{k\pm} = \pm q_k$ corresponding to the edges of the continuum band
into Eq.(\ref{E}) and get
\begin{equation}
\label{b}
U_{\rm bg}+U_{k\pm}=0
\;,
\end{equation}
which leads to the identity
$
  \left|
      b_k
  \right|
  = 1
$
and as a consequence to the violation of inequalities~(\ref{ineq}).
Eq.~(\ref{b}) allows to obtain the boundaries $\delta'_-$ and $\delta'_+$ of the interval
of $\delta'$ within which there is only one physical solution:
\begin{equation}
\label{dpm}
\delta'_\pm
=
\pm q_k
+
2 t_{\rm m}
\cos
\left(
    \pi k/k_{\rm L}
\right)
+
2 \tilde g^2/U_{\rm bg}
\,.
\end{equation}
For the values of parameters used in the numerical diagonalization, we find
$\delta'_-=-1.531\,\hbar\omega$
and
$\delta'_+=-1.479\,\hbar\omega$
in the case of attractive interaction, and
$\delta'_-=1.479\,\hbar\omega$
and
$\delta'_+=1.532\,\hbar\omega$
in the case of repulsive interaction.
The values of $\delta_-'$ are in perfect agreement with the results of numerical
calculations for $L=41$, while $\delta_+'$ have small deviations from the corresponding
numerical estimations.
In the special case $t_{\rm a}=t_{\rm m}=0$, $\delta'_-=\delta'_+=\delta'_*$ and Eq.~(\ref{b})
leads to the condition
\begin{equation}
\label{cond}
U_{\rm bg}
-
2
\tilde g^2/\delta'_*
=
0
\;.
\end{equation}
This is equivalent to the requirement that the effective scattering length
$
 a_{\rm bg}(1-\Delta B \Delta\mu/\delta')
$,
which appears in the mean-field theory as a result of the adiabatic elimination
of the molecular field~\cite{MVA}, vanishes.
The calculations presented above show that in the interval of the detunings
$\delta'_- < \delta' <\delta'_+$
the effective atomic interaction is gradually switched from the attractive
to the repulsive one.

The probabilities
$
\left|
    c_{jk}^{\rm a}
\right|^2
$,
$j=1,2,\dots$,
of the atomic states in Eq.(\ref{psi}) decrease with $j$ and have the form
\begin{eqnarray}
\left|
    c_{ik}^{\rm a}
\right|^2
=
\left(
    1 - b_k^2
\right)
b_k^{2(i-1)}
\left[
    1
    -
    \left|
        c_{0k}^{\rm a}
    \right|^2
    \left(
        1
	+
        S_k
    \right)
\right]
\;.
\end{eqnarray}
The soliton effective mass
\begin{equation}
m^*
=
\hbar^2
\left(
    \left.
    \frac
    {\partial^2 {\cal E}_k}
    {\partial k^2}
    \right|_{k=0}
\right)^{-1}
\\
=
\hbar^2
\left.
\frac
{
 {\cal E}_k
 +
 \left(
     U_{\rm bg}
     +
     U_k
 \right)
 S_k
}
{
 2
 t_{\rm m}
 \left(
     U_{\rm bg}
     +
     U_k
 \right)
 S_k
 - 4 t_{\rm a}^2
}
\right|_{k=0}
\end{equation}
is smaller than that at the bottom of the continuum
\begin{equation}
\tilde m
=
\hbar^2
\left(
    -
    \left.
    \frac
    {\partial^2 q_k}
    {\partial k^2}
    \right|_{k=0}
\right)^{-1}
=
\frac
{\hbar^2 k_{\rm L}^2}
{\pi^2 t_{\rm a}}
\;.
\end{equation}
The first two moments of the interatomic-distance distribution can be shown to be
\begin{eqnarray}
\langle
   w_k
\rangle
&=&
2
\left|
    c_{0k}^{\rm a}
\right|^2
b_k^2
/
\left[
    \left(
        1 - b_k^2
    \right)^2
    \left(
        1
	-
        \left|
            c_{k}^{\rm m}
        \right|^2
    \right)
\right]
\;,
\\
\langle
   w_k^2
\rangle
&=&
2
\left|
    c_{0k}^{\rm a}
\right|^2
b_k^2
\left(
    1 + b_k^2
\right)
/
\left[
    \left(
        1 - b_k^2
    \right)^3
    \left(
        1
	-
        \left|
            c_{k}^{\rm m}
        \right|^2
    \right)
\right]
\;.
\nonumber
\end{eqnarray}

In order to demonstrate the convergence to the limit $L\to\infty$,
we have plotted in Fig.~\ref{d} the quantities
$
  \Delta_{\rm s}
  =
  \sup_k
  \left|
      E_k^{({\rm s})}
      -
      {\cal E}_k
  \right|
  /
  \hbar\omega
$
as well as
$
  \Delta_{\rm t}
  =
  \sup_k
  \left|
      E_k^{({\rm t})}
      -
      q_k
  \right|
  /
  \hbar\omega
$
and
$
  \Delta_{\rm b}
  =
  \sup_k
  \left|
      E_k^{({\rm b})}
      +
      q_k
  \right|
  /
  \hbar\omega
$
for different $L$, where
$E_k^{({\rm s})}$, $E_k^{({\rm t})}$, and $E_k^{({\rm b})}$
are the eigenenergies of Eq.(\ref{evp}) corresponding to the soliton band,
the top and the bottom of the quasi-continuum band, respectively.
$\Delta_{\rm s}$ decreases exponentially with the increase of $L$ and it is very small
even for low values of $L$. The convergence for the boundaries of the continuum
band is slower, but the limit $L\to\infty$ describes quite well the results of
the numerical diagonalization already for a few tens of the lattice sites.
In addition, we have compared the results of the calculations
obtained on the basis of numerical solution
of the eigenvalue problem~(\ref{evp}) for $L=41$
which are presented in Figs.~\ref{s}-\ref{wdp}
with that worked out in the limit $L\to\infty$
and did not find any noticeable discrepancies.
This is consistent with the exponential decrease of $\Delta_{\rm s}(L)$.

In the absence of the magnetic field, the molecule creation is impossible
and one has to put $\tilde g=0$, $\delta'=0$, $t_{\rm m}=0$, in all the equations.
In this special case, the normalizable solution of Eq.~(\ref{E}) is given by
$
{\cal E}_k^{(0)}
=
{\rm sign}(U_{\rm bg})
\sqrt{
      U_{\rm bg}^2
      +
      q_k^2
     }
$,
which leads to the following expression for the effective mass
$
m^{*(0)}
=
-
\hbar^2
{\rm sign}(U_{\rm bg})
\sqrt{
      U_{\rm bg}^2
      +
      16 t_{\rm a}^2
     }
/
\left(
 4 t_{\rm a}^2
\right)
$.
These are exactly the results presented in Ref.~\cite{Scott}.
The soliton band exists again for repulsive as well as attractive atomic
interaction, but in the case of repulsive interaction it appears to be
a highly excited mode with the energy above the continuum band.

\section{Conclusion}

Summarizing, we have investigated quantum lattice solitons in a system of two
ultracold bosons near the Feshbach resonance.
Binding energy, effective mass, and spatial width of the solitons,
can be manipulated varying the detuning from
the Feshbach resonance.
In the case of attractive atomic interactions, the molecule creation stabilizes
the solitons increasing their effective mass as well as the binding energy
and decreasing the width.
In the case of repulsive interactions, the molecule creation leads
to the possibility of existence of bright solitons in some interval of detunings
analogous to the corresponding classical system.
The presence of quantum fluctuations leads to the fact that the interatomic distance
is a random quantity. Its standard deviation is even larger than the mean value.

The classical limit of the problem studied in the present work was considered
in~\cite{konotop}. Our results for the relative populations of the atomic
and molecular components are in agreement with the corresponding classical results.
In order to understand the transition from quantum to classical solitions
it is necessary to perform analogous calculations for higher number of atoms.
This can be done employing the same method as in the present study.
However, one has to keep in mind
that the dimension of the Hilbert space increases rapidly with the increase
of the particle number and the number of lattice sites.

\ack
This work was partly supported by the INTAS (Project No. 01-855) and SFB/TR 12.
K.V.K. would like to thank the University of Bath for kind hospitality.

\section*{References}


\end{document}